\documentclass[runningheads]{llncs}
\usepackage[T1]{fontenc}
\usepackage{graphicx}
\usepackage{todonotes}
\usepackage{booktabs}
\usepackage{multirow}
\usepackage{caption}
\usepackage{amsmath}
\usepackage{amssymb}
\usepackage[backend=biber,style=numeric]{biblatex}
\newcommand{\reg}{\textup{reg}}
\usepackage{units} %Weiteres Mathepaket mit \nicefrac{Zähler}{Nenner} Umgebung für Brüche
\usepackage{eurosym}%Eurozeichen
\usepackage{csquotes} %Anführungszeichen
\usepackage[nocomma, short]{optidef}
\usepackage{url}

\begin{document}
\title{Risk management in multi-objective portfolio optimization under uncertainty}
%
%\titlerunning{Abbreviated paper title}
% If the paper title is too long for the running head, you can set
% an abbreviated paper title here
%
\author{Yannick Becker\inst{1,2,3} \and
Pascal Halffmann\inst{1} \and
Anita Schöbel \inst{1,2}}
\authorrunning{Becker et al.}
% First names are abbreviated in the running head.
% If there are more than two authors, 'et al.' is used.
%
\institute{Fraunhofer Institute for Industrial Mathematics, Fraunhofer-Platz 1, 67663 Kaiserslautern, Germany,\\ \and
University of Kaiserslautern-Landau, Erwin-Schrödinger-Straße 52, 67663 Kaiserslautern, Germany,\\ \and
\email{yannick.b.becker@itwm.fraunhofer.de}}

\maketitle              % typeset the header of the contribution

\begin{abstract}
In portfolio optimization, decision makers face difficulties from uncertainties inherent in real-world scenarios. These uncertainties significantly influence portfolio outcomes in both classical and multi-objective Markowitz models. To address these challenges, our research explores the power of robust multi-objective optimization. Since portfolio managers frequently measure their solutions against benchmarks, we enhance the multi-objective min-regret robustness concept by incorporating these benchmark comparisons. 

This approach bridges the gap between theoretical models and real-world investment scenarios, offering portfolio managers more reliable and adaptable strategies for navigating market uncertainties. Our framework provides a more nuanced and practical approach to portfolio optimization under real-world conditions.

\keywords{Multi-objective optimization  \and Uncertainty \and Robustness \and Portfolio optimization.}
\end{abstract}

\section{Motivation}

Traditional portfolio optimization with a single objective function often falls short in addressing the multifaceted nature of real-world problems, leading to sub-optimal decision-making. Nowadays the \textcite{Markowitz1952} model is interpreted as a multi-objective optimization problem, where we optimize return and risk among others such as solvency ratio or transaction volume simultaneously as individual objective functions \cite{Daechert2021}. This has led to a significant improvement for the asset allocation in practice. However, decision makers still encounter challenges stemming from real-world uncertainties such as incomplete knowledge and parameter fluctuations that impact portfolio optimization. 

Financial applications typically use fixed input parameters, but these are influenced by market changes and thus cannot be constant \cite{Fenn2011}. To address uncertain parameters in portfolio optimization, we investigate how multi-objective robust optimization,  a rather new field with challenges in adapting single-objective robustness concepts \cite{Ide2015}, can not only enhance the decision support for portfolio optimization but also improve the risk management. 

Therefore, we generalize the multi-objective adaption of the min-regret robustness by \textcite{Groetzner2022} by introducing so-called \enquote{benchmarks} in the min-regret-formulation to provide a more practical robustness concept \cite{Simoes2018}. Further, this allows for non-uncertain objective functions that may occur as well. We provide four methods to identify benchmarks for multi-objective portfolio optimization with uncertain risk objective and test these approaches on real data. This demonstrates the capacity of multi-objective optimization as a potent decision support tool for portfolio optimization.

%Different scenarios with unknown probability, such as financial crises or economic growth, are considered, influencing only the risk parameters. 
%To mitigate this uncertainty within a single-objective function, we present an innovative multi-objective robust optimization concept, extending the established min-regret-robustness from single-objective optimization. 

%The findings of this study demonstrate the capacity of multi-objective optimization to provide a potent decision support system for portfolio optimization. This empowers decision-makers grappling with incomplete and uncertain data to make more informed and robust asset allocation decisions, proving particularly relevant in dynamic and uncertain financial markets.

%\section{Multi-objective portfolio optimization}
Consider $n$ tradeable assets with expected returns $\mu_{i}$ and covariances $\sigma_{ij}$ for $i,j=1,...,n$. We consider the expected portfolio return and variance in line with \textcite{Korn2001} as $f_{\mu}(x) = x^{T}\mu$ and $f_{\sigma} = x^{T}\sigma x$, where $\mu=(\mu_{1},...,\mu_{n})^{T}$ and $\sigma=(\sigma_{ij})_{i,j=1,..,n}$. Since we aim to optimize all objectives simultaneously, we state the \emph{Mean-Variance} problem as
\normalsize
\begin{mini*}[2]
{x\in \Delta^{n}}{ \left( \begin{array}{c} - f_{\mu}(x) \\ f_{\sigma}(x) \end{array}	\right),}{}{(MV)\quad}
\end{mini*}
\normalsize
where $\Delta^{n} = \{ x = (x_{1}, \ldots, x_{n})^{T} \in \mathbb{R}^{n} \mid \sum_{i=1}^{n} x_{i} = 1 \text{ and } x_{i} \geq 0 \, \text{ for } \, i = 1, \ldots, n \}$ is the standard n-simplex. Multi-objective optimization endeavors to navigate the trade-offs between two or more conflicting objectives and therefore leads to a Pareto front, i.e. a spectrum of efficient solutions each offering a different balance of trade-offs among the objectives. We refer to \textcite{Halffmann2022} for an comprehensive introduction to multi-objective optimization and its notation. 
%In order to calculate this front we transform $(MV)$ using the $\epsilon$-constraint scalarization technique and receive a parametrized family of optimization problems
%\begin{align*}
%	(MV(\epsilon), \epsilon \in [\mu_{min},\mu_{max}]) &&\text{with} && MV(\epsilon): & \quad \min \quad f_{\sigma}(x) \\
%	&&&&& \quad \text{ s.t.}\ \quad f_{\mu}(x)\leq\epsilon \\
%	&&&&& \quad \quad \quad \quad x \in \Delta^{n}
%\end{align*}
%where $\mu_{min} = \underset{i=1,...,n}{\min} \mu_{i}$ and $\mu_{max} = \underset{i=1,...,n}{\max} \mu_{i}$ are the minimum and maximum expected return values.

\section{Multi-objective \textit{robust} optimization in general}
A weighted objective function or one objective function for each scenario is not practical for multi-objective problems \cite{Ehrgott2014}. Hence, an uncertain optimization problem incorporates some given uncertainty set $\mathcal{U}\subseteq\mathbb{R}^{m}$ by parameterizing a given problem $(P)$ in $\mathcal{U}$. Thus, we have $(P(\xi), \xi\in\mathcal{U})$ with $P(\xi): \min_{x \in \mathcal{X}} f(x, \xi)$, where $f:\mathbb{R}^{n} \times \mathcal{U} \rightarrow \mathbb{R}^{k}$.
A common technique to approach an uncertain problem is to formulate a well-defined robust counterpart \cite{BenTal1998}, which defines what a robust solution is. With respect to the practical application of portfolio selection, portfolios are compared against the best competitors \cite{Simoes2018}. Therefore, the concept of regret robustness seems to be appropriate. In regret robustness one compares a solution $x$ with what could have been achieved if the scenario was known before (in the worst case over all scenarios). The regret robust counterpart has recently been adapted to multi-objective optimization by \textcite{Groetzner2022}:

\normalsize
\begin{mini*}[2]
	{x\in \mathcal{X}}{ \reg(x, \mathcal{U}),}{}{(rRC)\quad}
\end{mini*}
\normalsize
where we set $\reg(x, \mathcal{U}) = (\reg_{i}(x, \mathcal{U}))_{i=1,...,n}$ and $\reg_{i}(x, \mathcal{U}) = \underset{\xi \in \mathcal{U}}{\max}\,\reg_{i}(x, \xi)
		= \underset{\xi \in \mathcal{U}}{\max} (f_{i}(x, \xi) - f_{i}^{*}(\xi))$, with
 $f_{i}:\mathbb{R} \times \mathcal{U} \rightarrow \mathbb{R}, i=1,...,n$, and $f_{i}^{*}(\xi) = \underset{b\in\mathcal{X}}{\min} \, f_{i}(b,\xi)$ is the optimal solution for minimizing $f_{i}$ if the scenario $\xi$ was known. 

But what if we want to compare our solution not with $f_{i}^{*}(\xi)$? In practice, portfolios of asset managers are commonly compared against certain benchmarks pressuring them to not fall too short of such particular benchmarks. \textcite{Simoes2018} investigated the robust regret for single-objective optimization under the use of benchmarks. I.e. for an uncertain objective function $f:\mathbb{R}^{n} \times \mathcal{U} \rightarrow \mathbb{R}$, an uncertainty set $\mathcal{U}$ and a set of benchmarks $\mathcal{B}=\{b_{1},...,b_{m}\}\subseteq\mathbb{R}^{n}$ we consider $	\reg(x,\mathcal{U},\mathcal{B}) = \underset{\xi\in\mathcal{U}}{\max}\, \reg(x,\xi, \mathcal{B})
	 = \underset{\xi\in\mathcal{U}}{\max}\, (f(x,\xi) - \underset{b\in\mathcal{B}}{\min}f(b,\xi))$.

\noindent\textbf{Extension with benchmarks:} Extending the approach in \cite{Groetzner2022} by using benchmarks, we obtain a \textit{multi-objective robust optimization with benchmarks} problem:
%As already discussed, asset managers nowadays aim to consider uncertainties, optimize multiple criteria simultaneously, and find a solution within certain benchmark areas. Therefore, we extend the approach of \textcite{Groetzner2022} by using benchmarks and obtain a \textit{multi-objective robust optimization with benchmarks} problem:

\normalsize
\begin{mini*}[2]
	{x\in \mathcal{X}}{ \reg(x, \mathcal{U}, \mathcal{B}).}{}{(rRCb)\quad}
\end{mini*}
\normalsize
\section{Multi-objective partially robust optimization with benchmarks}
In our work, we consider uncertainty only in one objective function and investigate a \textit{multi-objective partially robust optimization with benchmarks} problem:

\normalsize
\begin{mini*}[2]
	{x\in \mathcal{X}}{  \left( \begin{array}{c} f_{1}(x) \\ \vdots \\ f_{n-1}(x) \\ \underset{\xi\in\mathcal{U}}{\max}\, (f_{n}(x,\xi) - f_{n}(b^{\xi}, \xi)) \end{array}	\right),}{}{(prRCb)\quad}
\end{mini*}
\normalsize
where $f_{n}$ is the uncertain objective and $b^{\xi} \in \operatorname*{arg\,min} \{f_{n}(b,\xi) \, | \, b\in\mathcal{B}\}$ the pre-calculated benchmarks for scenario $\xi\in\mathcal{U}$.  %$\displaystyle b^{\xi} \in \operatorname*{arg\,min}_{b\in\mathcal{B}} f_{n}(b,\xi)$

\subsection{Adaption to portfolio optimization}
Since benchmarks are highly context specific, we now adapt our approach to portfolios in finance. We assume the covariance to be uncertain so that we have uncertainty only in $f_{\sigma}$ of $(MV)$. Motivated by practice, we consider only three different scenarios, denoted as \textit{regimes} in finance, namely \textit{Crisis (C)}, \textit{Normality (N)}, and \textit{Growth (G)} which leads to the uncertainty set $\mathcal{U} = \{\sigma^{C}, \sigma^{N}, \sigma^{G} \}$. This allow us to simplify $(rRCb)$ to a two-stage problem by first calculating one benchmark $b^{\sigma}$ for each $\sigma \in \mathcal{U}$ and then solving the bi-objective problem:

\normalsize
\begin{mini*}[2]
	{x\in \Delta^{n}}{   \left( \begin{array}{c} - \mu^{T} x \\ \underset{\sigma\in\mathcal{U}}{\max}\, |x^{T}\sigma x - (b^{\sigma})^{T} \sigma b^{\sigma}| \end{array}	\right),}{}{(prRCb^{*})\quad}
\end{mini*}
\normalsize
where $b^{\sigma}, \sigma \in\mathcal{U}$, is a $\sigma$-unique benchmark. We have chosen the absolute regret since we aim to obtain a robust Pareto front with results lower and greater in variance from $b^{\sigma}$. Further, we identify efficient solutions $x^*$ for the benchmarks $b^{\sigma}=f_\sigma(x^*)$ without involving decision makers by one of the following four techniques that are well-used in practice to find suitable portfolios:\newpage

\begin{enumerate}
	\item[(i)] \textbf{Bounded risk:} $\displaystyle x^*\in\operatorname*{arg\,max}_{\{x\in\Delta{^{n}}| f_{\sigma}(x) =0.03\}} f_\mu(x)$.
	\item[(ii)] \textbf{Weighted sum with risk aversion:} $\displaystyle x^*\in\operatorname*{arg\,min}_{x\in\Delta{^{n}}}\left( 3.5 \cdot f_{\sigma}(x) + f_{\mu}(x) \right)$.
	\item[(iii)] \textbf{Sharpe ratio:} $\displaystyle x^*\in\operatorname*{arg\,max}_{x\in\Delta{^{n}}}\frac{f_{\mu}(x)}{f_{\sigma}(x)}$.
	\item[(iv)] \textbf{Inner 60\% percentile in risk:} 
 $\displaystyle x^*\in\operatorname*{arg\,max}_{\{x\in\Delta{^{n}}| f_{\sigma}(x)-\sigma_{min} \leq\nicefrac{1}{5}(\sigma_{max} - \sigma_{min})\}} f_\mu(x)$.
\end{enumerate}

\subsection{An illustrative example}
\begin{figure}[tb]
    \centering
	\includegraphics[width=0.69\linewidth]{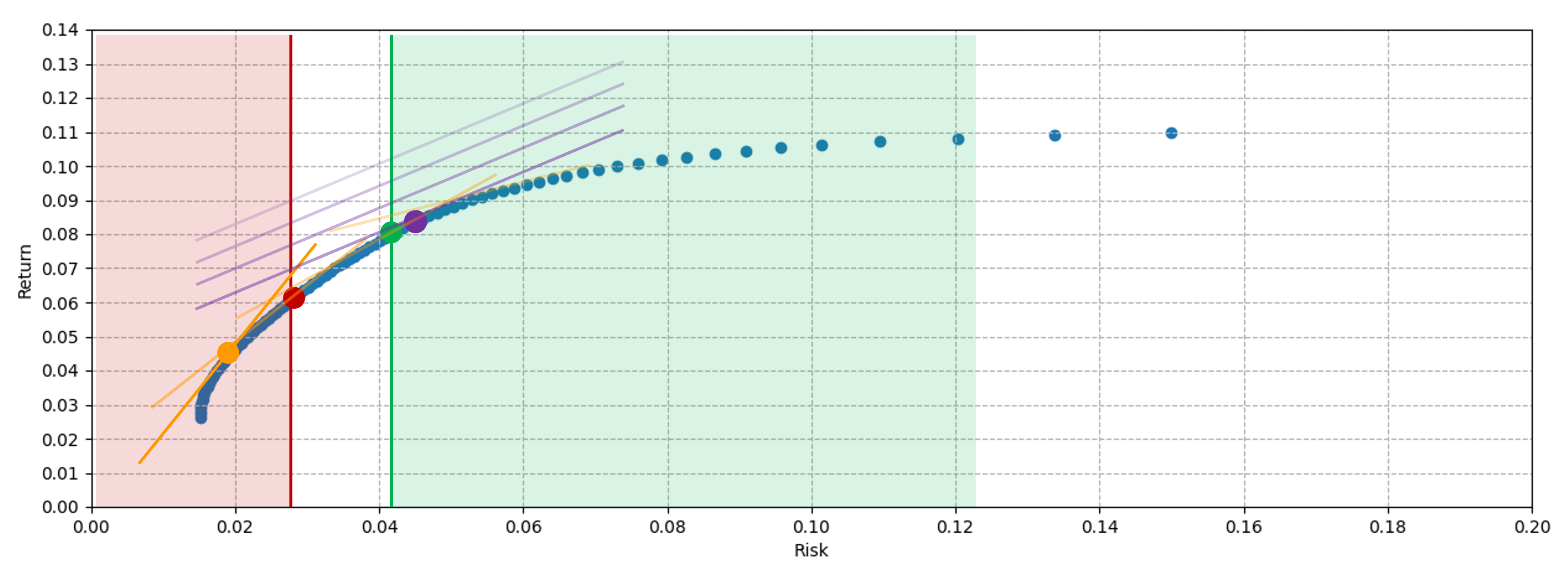}
	\caption{Pareto front and benchmarks (bounded risk (red), weighted sum (purple), Sharpe ratio (orange), and 60\% percentile (green)) for regime C.}
	\label{fig:bench}
\end{figure}

\begin{figure}[tb]
    \centering
	\includegraphics[width=0.69\linewidth, angle=90]{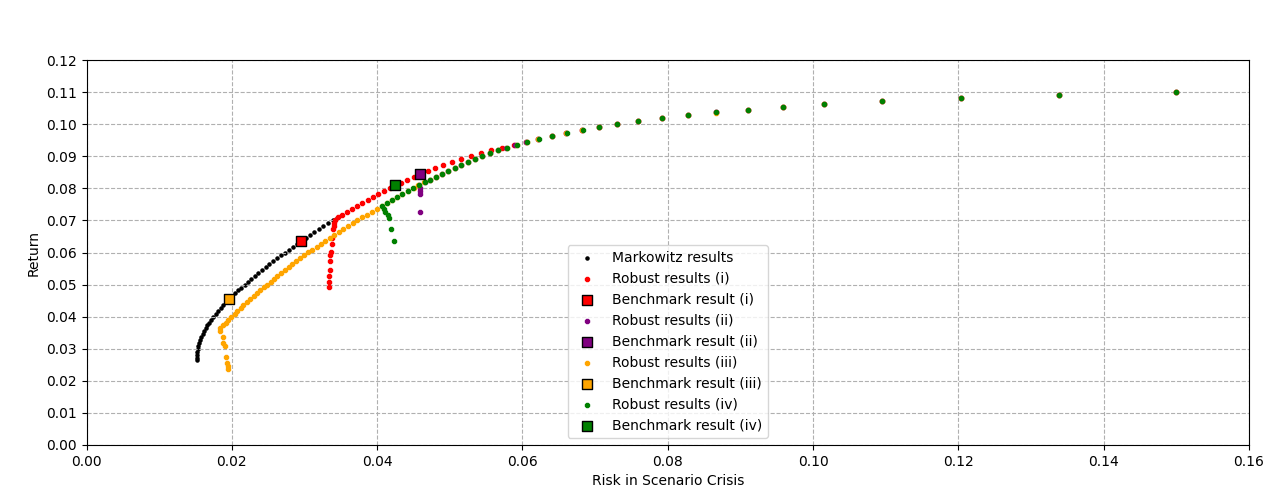}
    \includegraphics[width=0.69\linewidth, angle=90]{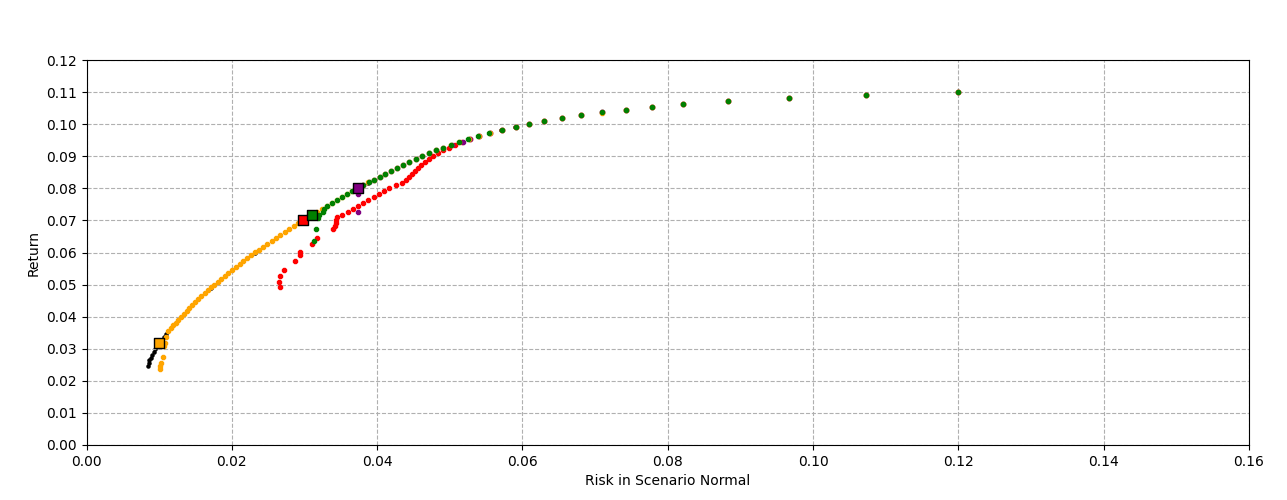}
    \includegraphics[width=0.69\linewidth, angle=90]{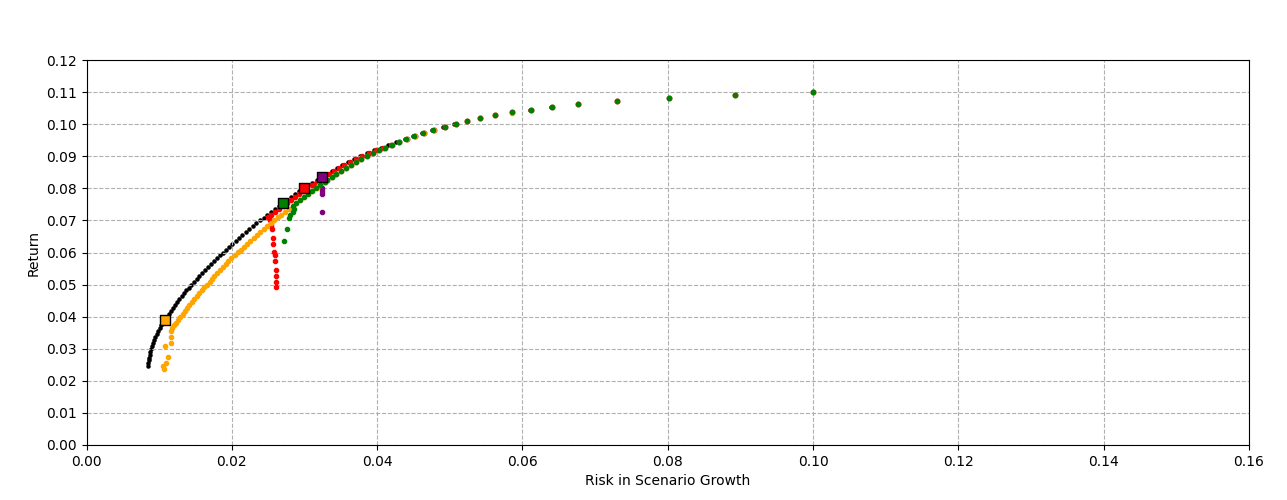}
	\caption{Results for $(prRCb^{*})$ on sample asset data: for every benchmarks approach, the robust results' risk are transformed to specific risks and the sets of robust solutions are compared to the actual Pareto front for each scenario.}
	\label{fig:robust-crisis}
\end{figure}

We consider 15 different asset classes with expected returns ranging from 2\% to 11\% and standard deviations ranging from 1\% to 15\%. For each of the three different regimes (\textit{C}, \textit{N}, \textit{G}), we have specific correlation matrices and standard deviations as detailed in \cite{beckerya_24}. In Regime C, correlations are higher, indicating stronger co-movements between asset classes. Regime N features moderate correlations, while regime G shows lower correlations, indicating a higher effectiveness of diversification. First, we compute the benchmarks based on each of the described techniques (i) - (iv) (see Fig. \ref{fig:bench} for regime C). Next, we obtain the Pareto front of $(prRCb^{*})$ for each benchmark approach, Figure \ref{fig:robust-crisis}. 
When analyzing these results, we immediately see that there are barely any robust solutions left of the benchmarks. Obviously, this is due to the existence of portfolios with higher return and similar regret regarding the risk (distance to the risk of the benchmark portfolio) dominating solutions left of the benchmarks. Conversely, it seems that high risk/return robust-optimal solutions are Pareto-optimal for every scenario. As there are only few assets and thus not many portfolios with such a high risk/return, fluctuations between scenarios cannot be mitigated by changing portfolio weights. The interesting area for asset managers is around the benchmarks, where we observe an increasing distance between robust optimal solutions and the Pareto fronts. Many feasible portfolios exist and fluctuations between scenarios have a high impact. As we aim to minimize the regret towards the benchmarks, this results in sacrificing return for robustness in risk. At last, extreme benchmark solutions, e.g. the ones we obtain by Sharpe ratio and weighted sum, are not very useful in practice as they either have a similar effect as a worst-case robustness or only provide solutions with high risk.
%Note that we commonly aim to minimize risk, but for the sharpe ratio we additionally consider the expected return which is ment to be maximized. 
%Table \ref{tab:benchmarks} summarizes the benchmarks $b^{\sigma}, \sigma \in \mathcal{U}$, for four different strategies across three regimes. Using these benchmarks we now get to the robust optimization problem $(prRCb^{*})$.
%\begin{table}[h!]  % Platzierungsoption [h!] für stark bevorzugte Platzierung hier
%	\centering
%	\caption{Benchmarks for different strategies across three market regimes.}
% 	\begin{tabular*}{\textwidth}{@{\extracolsep{\fill}} l|cccc}
%		\toprule
%		Regime & Bounded risk & Weighted sum & Sharpe ratio & 60\% percentile \\
%		\midrule
%		Crisis & .0636 (.0296) & .0845 (.0460) & .0455 (.0196) & .0809 (.0425) \\
%		Normal & .0700 (.0298) & .0800 (.0373) & .0318 (.0100) & .0718 (.0311) \\
%		Growth & .0800 (.0299) & .0836 (.0324) & .0391 (.0108) & .0755 (.0270) \\
%		\bottomrule
%	\end{tabular*}	
%	\label{tab:benchmarks}
%\end{table}

\section{Discussion and Outlook}
So, which benchmarks approach works best? An answer requires a quantitative concept for the evaluation, which is challenging, since balancing between robustness over all scenarios and single-scenario performance is a multi-objective problem itself. A first approach comparing different concepts is introduced by \textcite{Schoebel2021}, where robust solution sets are evaluated regarding a nominal (e.g. most probable) and a worst-case scenario. While a nominal scenario (regime $N$) could make sense for our use case, a worst-case evaluation contradicts our min-regret approach. Essentially, a metric that evaluates the performance of robustness concepts could be a robustness concept itself, so why not using this metric instead? One reason is that some metrics can only be applied a-posteriori or utilize the Pareto front of every scenario; for example the worst-case hypervolume ratio between scenario Pareto fronts and robust solution set. Clearly, Sharpe ratio with rather extreme benchmarks performed best. From a practical point of view, bounded risk and 60\% percentile provide suitable portfolios with the former in favour, due to a better spread around the benchmarks. This evaluation of robustness concepts is worth investigating further. Future research may also focus on adding additional objective functions like solvency, uncertainty in another objective function, different uncertainty sets, or a combination of these.
\begin{credits}
\subsubsection{\ackname} We would like to thank Ralf Korn for the valuable discussions on finding benchmarks in portfolio optimization.
\subsubsection{\discintname}
The authors have no competing interests to declare that are relevant to the content of this article.
\end{credits}
\printbibliography
\end{document}